\documentclass[sigconf]{acmart}
\usepackage{amsmath,amsfonts,bm}

\def\eqref#1{equation~\ref{#1}}
\def\ceil#1{\lceil #1 \rceil}

\def\1{\bm{1}}

\DeclareMathAlphabet{\mathsfit}{\encodingdefault}{\sfdefault}{m}{sl}
\SetMathAlphabet{\mathsfit}{bold}{\encodingdefault}{\sfdefault}{bx}{n}

\DeclareMathOperator*{\argmax}{arg\,max}

\usepackage[inline]{enumitem}
\usepackage{booktabs} %
\usepackage{hyperref}
\usepackage{url}
\usepackage{array}
\newcommand{\PreserveBackslash}[1]{\let\temp=\\#1\let\\=\temp}
\newcolumntype{C}[1]{>{\PreserveBackslash\centering}p{#1}}
\newcolumntype{R}[1]{>{\PreserveBackslash\raggedleft}p{#1}}
\newcolumntype{L}[1]{>{\PreserveBackslash\raggedright}p{#1}}
\usepackage{multirow}
\usepackage{colortbl}
\usepackage{graphicx}
\usepackage{pifont}
\usepackage{xspace}
\usepackage{subcaption}
\usepackage{balance}

\copyrightyear{2021} 
\acmYear{2021} 
\setcopyright{acmlicensed}\acmConference[SIGIR '21]{Proceedings of the 44th International ACM SIGIR Conference on Research and Development in Information Retrieval}{July 11--15, 2021}{Virtual Event, Canada}
\acmBooktitle{Proceedings of the 44th International ACM SIGIR Conference on Research and Development in Information Retrieval (SIGIR '21), July 11--15, 2021, Virtual Event, Canada}
\acmPrice{15.00}
\acmDOI{10.1145/3404835.3462889}
\acmISBN{978-1-4503-8037-9/21/07}

\newcommand{\tsc}[1]{\textsuperscript{#1}} %shorthand for superscripts
\author{Sebastian Hofst{\"a}tter\tsc{1}, Bhaskar Mitra\tsc{2}, Hamed Zamani\tsc{3}, Nick Craswell\tsc{2}, Allan Hanbury\tsc{1}}
\affiliation{
  \institution{\vskip .2cm}%add some spacing if needed
  \institution{\tsc{1} TU Wien, \tsc{2} Microsoft , \tsc{3} University of Massachusetts Amherst}
}
\affiliation{%
 \institution{\vskip .1cm}%add some spacing if needed
  \institution{$^{1}$ s.hofstaetter@tuwien.ac.at, hanbury@ifs.tuwien.ac.at,  $^{2}$ \{bmitra, nickcr\}@microsoft.com, $^{3}$ zamani@cs.umass.edu}
}

\renewcommand{\authors}{Sebastian Hofst{\"a}tter, Bhaskar Mitra, Hamed Zamani, Nick Craswell, Allan Hanbury}

\newcommand{\eg}{\emph{e.g.}}
\newcommand{\wrt}{\emph{w.r.t. }}

\newcommand{\model}{IDCM\xspace}
\newcommand{\efficientmodel}{ESM\xspace}
\newcommand{\effectivemodel}{ETM\xspace}

\title{Intra-Document Cascading: Learning to Select Passages for~Neural~Document~Ranking}

\begin{abstract}
An emerging recipe for achieving state-of-the-art effectiveness in neural document re-ranking involves utilizing large pre-trained language models---\eg, BERT---to evaluate all individual passages in the document and then aggregating the outputs by pooling or additional Transformer layers.
A major drawback of this approach is high query latency due to the cost of evaluating every passage in the document with BERT.
To make matters worse, this high inference cost and latency varies based on the length of the document, with longer documents requiring more time and computation.
To address this challenge, we adopt an \emph{intra-document cascading} strategy, which prunes passages of a candidate document using a less expensive model, called ESM, before running a scoring model that is more expensive and effective, called ETM. 
We found it best to train ESM (short for Efficient Student Model) via knowledge distillation from the ETM (short for Effective Teacher Model) e.g., BERT.
This pruning allows us to only run the ETM model on a smaller set of passages whose size does not vary by document length.
Our experiments on the MS MARCO and TREC Deep Learning Track benchmarks suggest that the proposed Intra-Document Cascaded Ranking Model (\model) leads to over $400\%$ lower query latency by providing essentially the same effectiveness as the state-of-the-art BERT-based document ranking models.

\end{abstract}

\DeclareMathOperator*{\bert}{BERT}
\DeclareMathOperator*{\ck}{CK}
\DeclareMathOperator*{\cnn}{CNN}

\begin{document}

\maketitle

\section{Introduction}
\label{sec:intro}

Ranking documents in response to a query is a core problem in information retrieval (IR). Many systems incorporate ranking as a core component, such as Web search engines and news search systems.
Other systems build on document ranking, e.g., an agent capable of conversation and disambiguation may still have document ranking as a component~\cite{zamani2020macaw}. Therefore, retrieval model improvements are likely to lead to ``lifting all boats''. 

One difficulty in document ranking is that documents can vary significantly in length.
In traditional IR, variation in length can be explained by (1) the verbosity hypothesis, that the author used more words to explain the topic, or (2) the scope hypothesis, that the author covered multiple topics \cite{robertson2009probabilistic}.
In practice, each hypothesis is a partial explanation for document length, and retrieval models typically apply document length normalization.
In other words, a document with some irrelevant parts can still be considered relevant overall, because having somewhat broader scope than the current query does not rule out a document from being a useful result.

One way to deal with documents of varying length with some irrelevant parts is to use passage-level evidence for document ranking.
Early studies \cite{callan1994passage, kaszkiel1997passage} found that a passage of text need not be defined based on document structure, such as paragraphs or sentences.
A good approach was to  divide the content into fixed-size windows of 150 words or more, compare the query to all passages, then score the document based on the score of its highest-scoring passage.
This is consistent with the scope hypothesis, that we should focus on finding some relevant content, without penalizing the document for having some irrelevant content.

In neural IR, it is possible to significantly outperform classic retrieval systems \cite{craswell2019overview, lin2020pretrained}.
This is often done by taking multiple fixed-size windows of text, applying a deep neural network to score each passage, and scoring the document based on the highest-scoring passages.
This is similar to the classic IR approaches in \cite{callan1994passage, kaszkiel1997passage}, but works much better as the per-passage models are more effective.

However, the problem with the neural approaches is the cost of applying the per-passage model. Applying neural net inference for every passage in every document being ranked requires significant computation and leads to higher query latency. This limits the impact of the new neural approaches, since if the cost of inference is too high, it cannot be used in large-scale production systems. To avoid this problem, some models retrieve documents solely based on their first passage~\cite{craswell2019overview}, however, this is a sub-optimal solution.

In this work, we address this issue by proposing an Intra-Document Cascading Model (\model) that employs an in-document cascading mechanism with a fast selection module and a slower effective scoring module.\footnote{In our experiments, we used DistilBERT~\cite{sanh2019distilbert}, an efficient variant of BERT as the ``slow effective scoring module''. Throughout this paper, we refer to it as the BERT model.} This simultaneously provides lower query latency and state-of-the-art ranking effectiveness.
We evaluate our model on two document ranking query sets:
\begin{enumerate*}[label=(\roman*)]
    \item TREC DL 2019~\citep{craswell2019overview},
    \item MSMARCO~\citep{msmarco16}. 
\end{enumerate*}
We study how to train the \model architecture in multiple stages to control the collaboration of the cascading sub-modules, and investigate: 
\newcommand{\RQone}{\begin{itemize}
    \item[\textbf{RQ1}] Can \model achieve comparable effectiveness to the full BERT-based ranker at lower computation cost and query latency?
\end{itemize}}
\RQone
Among the different variants we explore in this work, we found that training the selection module using \emph{knowledge distillation} based on passage-level labels derived from the BERT score in conjunction with an intra-document ranking loss achieves the best overall re-ranking quality.
Under this setting, the selection module is trained to approximate the passage ranking that would have been produced if ordered by their corresponding BERT-scores, to filter out non-relevant parts of the document.

An important hyperparameter in this setting is the number of passages per document that survives the selection stage for the subsequent BERT-based evaluation.
Here we study:
\newcommand{\RQtwo}{\begin{itemize}
    \item[\textbf{RQ2}] How is the effectiveness-efficiency trade-off influenced by the number of passages $k$ that the less expensive model selects from the document?
\end{itemize}}
\RQtwo

In our baseline setting, the BERT model inspects up to the first $k$ passages.
We observe superior performance with $k=40$ compared to smaller values of $k$.
However, the proposed \model framework achieves roughly the same ranking effectiveness by applying the BERT model to only the top $4$ passages pre-selected by the less-expensive preceding model. 
Consequently, this result comes at a much lower query latency.

The application of BERT to multiple in-document passages has the undesirable property of introducing large variance in query response time depending on the length of candidate documents.
Under the \model settings, this variance is largely reduced because the expensive BERT model is applied to $k$ top passages for every document, unless the document is so short that it has less than $k$ passages.
This leads us to our next research question:

\newcommand{\RQthree}{\begin{itemize}
    \item[\textbf{RQ3}] How does \model compare to the baseline with respect to variance in query latency?
\end{itemize}}
\RQthree

We observe that our baseline setup of using BERT over all passages has a very high standard deviation and long tail \wrt query latency, whereas \model has much more predictable query latency centered around the mean.
This is partly because our passage selection module is fast enough to triage up to $40$ candidate passages in the same time as the BERT model takes to evaluate a single passage.
Therefore, while the contribution of the selection stage to query latency is still a function of document length, the variance is largely reduced.

Finally, we study the how the passage selection under \model compares to the highest-scoring passages by the BERT model.

\newcommand{\RQfour}{\begin{itemize}
    \item[\textbf{RQ4}] How often does the passage selection under \model recall the same passages as those scored highly by the BERT model?
\end{itemize}}
\RQfour

Overall, we observe that the selection module recalls $60$-$85\%$ of the top BERT-scored passages depending on the value of $k$.
We also find that passages closer to the beginning of the document are more likely to be selected by both models, although there is a long tail of passage selection from lower positions within the document.
Interestingly, we also observe that the selection module achieves a better recall in the case of relevant documents, which may be explained by the relevant passages being easier to distinguish from the nonrelevant passages in a relevant document, compared to the same in case of a nonrelevant document.

To summarize, this paper proposes a cascaded architecture and training regime to address the high and variable query latency associated with applying BERT to evaluate all in-document passages for the document ranking task.
We demonstrate that employing our approach leads to lower mean and variance for query latency, and enables broader scope of application for BERT-based ranking models in real-world retrieval systems.

\begin{itemize}[leftmargin=*]
    \item We propose \model, an intra-document cascade ranking model, including a training workflow using knowledge distillation.
    \item We evaluate our approach on TREC-DL'19 and MSMARCO DEV; and show that \model achieves similar effectiveness on average at four times lower latency compared to the BERT-only ranking model.
    \item We perform extensive ablation studies to validate our multi-stage training approach and the benefits of knowledge distillation for optimizing the selection module of \model.
\end{itemize}

To improve the reproducibility of our work, we open-source our implementation and release the trained models at: \\ \url{https://github.com/sebastian-hofstaetter/intra-document-cascade}

\section{Related Work}
Since demonstrating impressive performance on passage ranking~\citep{nogueira2019passage}, several successful efforts are underway to employ BERT~\citep{devlin2018bert} and other Transformer~\citep{vaswani2017attention} based models to the document ranking task~\citep{Hofstaetter2020_sigir,craswell2019overview}.
However, the quadratic memory complexity of the self-attention layer with respect to the input sequence length poses a unique challenge in the way of scaling these models to documents that may contain thousands of terms.
While some efforts are underway to directly address the scalability of the self-attention layer in these ranking models~\citep{mitra2020conformer,beltagy2020longformer,ainslie2020etc}, a majority of applications of Transformer-based architectures to document ranking involves segmenting the document text into smaller chunks of text that can be processed efficiently~\citep{yan2019idst, Hofstaetter2020_sigir, li2020parade}.

\begin{figure*}[t]
    \centering
    %trim={<left> <lower> <right> <upper>}
    \includegraphics[width=0.85\textwidth,clip, trim=0.2cm 0.2cm 0.2cm 0.2cm]{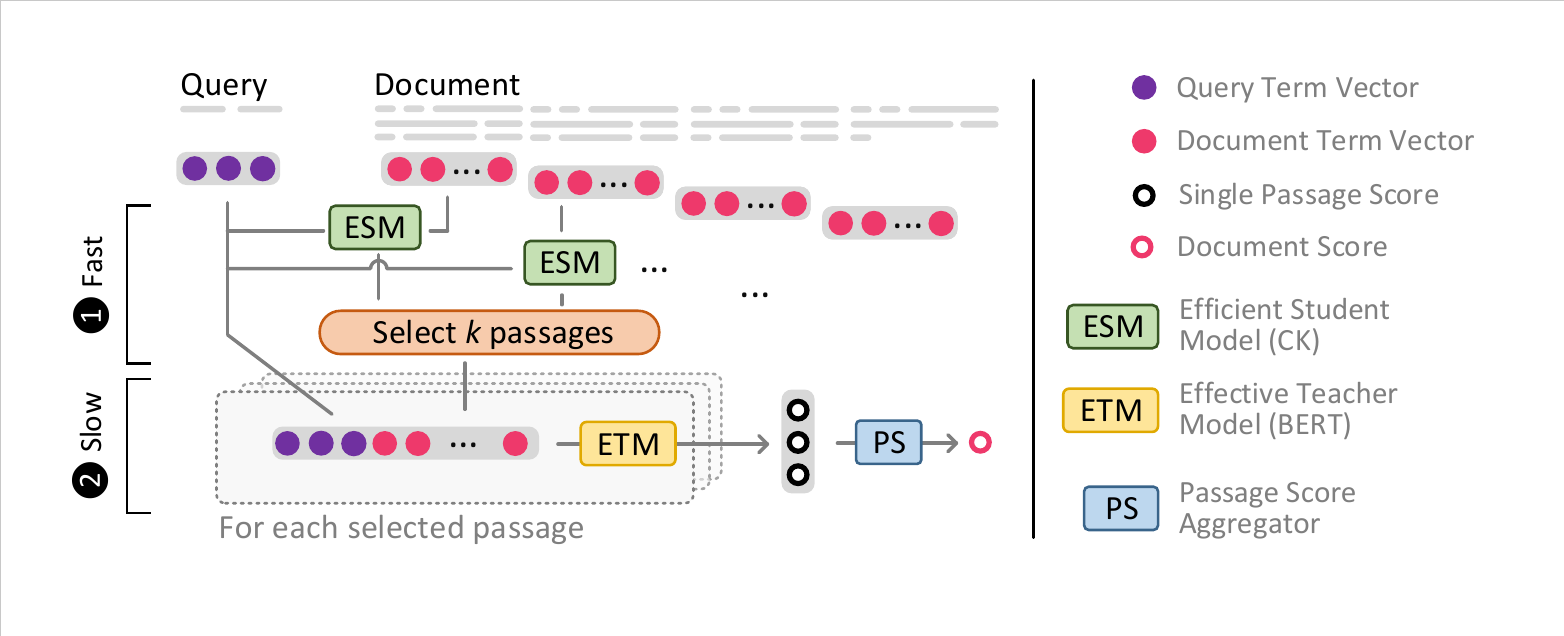}
    \vspace{-0.6cm}
    \caption{The \model architecture that consists of two cascading stages: \ding{202} To allow for long-document input, the first stage is a lightweight and fast selection model. \ding{203} Only the top k passages from the selection model are scored with a costly BERT-based scoring module to form the final document score.}
    \label{fig:model-workflow-combined}
    \vspace{-0.2cm}
\end{figure*}

The idea of using passage-level evidence for document ranking is not unique to neural methods.
Several classical probabilistic~\citep{callan1994passage} and language model~\citep{liu2002passage, bendersky2008utilizing} based retrieval methods---as well as machine learning based approaches~\citep{sheetrit2020passage}---incorporate passage-based relevance signals for document ranking.
An underlying assumption across many of these approaches is that all passages from the document are inspected by the model.
However, as our models become more computation and memory intensive, it quickly becomes expensive and even infeasible to evaluate every passage from the document.
This is exactly where text selection using light-weight approaches becomes necessary so that the more complex model only has to inspect the most important parts of the document, which is the main motivation for this work.

One could draw parallels between our approach of using cheaper models to detect interesting regions of the document to the highly influential work of \citet{viola2001rapid} in computer vision for fast object detection using cascades of models of increasing complexity.
Similarly, cascaded approaches have also been employed extensively~\citep{wang2011cascade, chen2017efficient, gallagher2019joint, nogueira2019multi} in IR to progressively rank-and-prune the set of candidate documents, from the full collection all the way to the final ten or so results presented to the user.
Unlike these approaches that employ a cascade of models to prune the candidate set of documents, we use the cascaded approach to prune the set of regions within the document that needs to be inspected by the expensive and effective ranking model (i.e., intra-document cascading).

In a cascaded setting, typically the models are trained progressively, starting with the simplest model that is exposed to the largest number of candidates, and then subsequent models are trained using a data distribution that reflects the candidates that survive the earlier stages of pruning.
This sequential training strategy is sometimes referred to as telescoping~\citep{matveeva2006high}.
Joint optimization of the different rankers within a multi-stage cascaded search architecture has also been explored~\citep{gallagher2019joint}.

In our work, we adopt a different training strategy.
We begin by training the costly ranking model and then apply knowledge distillation to train the model for the preceding selection stage.
Our approach is motivated by the strong empirical performance observed in recent work exploring knowledge distillation from larger BERT to more effective models~\citep{sanh2019distilbert, jiao2019tinybert, tang2019distilling, hofstaetter2020_crossarchitecture_kd}.

\section{The Intra-Document Cascade Model}
\label{sec:coral}
Neural ranking models have resulted in significant improvements in a wide variety of information retrieval tasks. However, there exists an efficiency-effectiveness trade-off in these models. In other words, state-of-the-art neural ranking models mostly suffer from high query latency and high GPU memory requirements. A popular solution to address the GPU memory issue is to divide documents into multiple passages and compute the retrieval scores for each passage \cite{yilmaz2019cross,yan2019idst,nogueira2019document}. For instance, \citet{dai2019deeper} compare considering only the first passage of the document with scoring every passage in the document and use the highest passage score as the document score. We believe that the existing approaches lead to either sub-optimal ranking or high computational cost. 

Inspired by previous work on passage-level evidences for document retrieval~\cite{liu2002passage,callan1994passage}, in this section we propose an alternative efficient and effective solution by introducing the Intra-Document Cascade Model named \textbf{\model}. In the following, we first introduce the \model architecture and optimization, and then describe its implementation details.

\begin{figure*}[t]
    \centering
    %trim={<left> <lower> <right> <upper>}
    \includegraphics[width=0.75\textwidth,clip, trim=0.1cm 0.1cm 0.1cm 0.1cm]{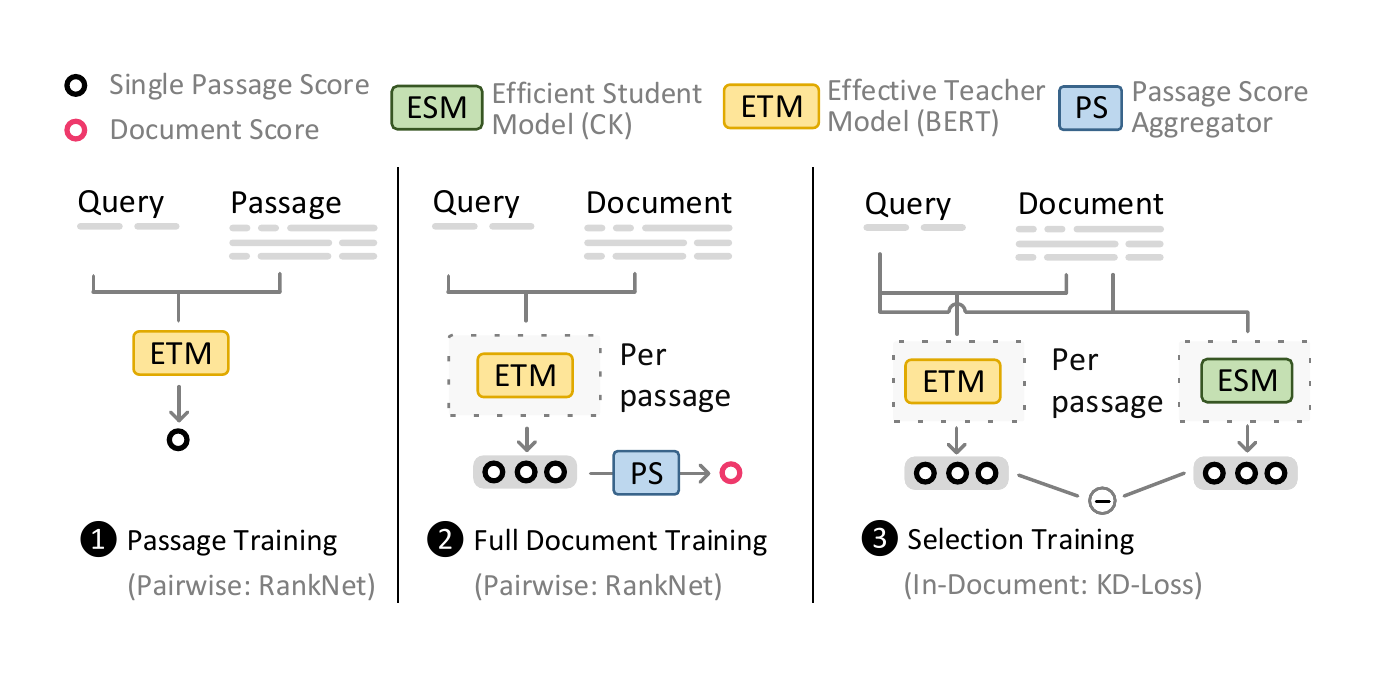}
    \vspace{-0.7cm}
    \caption{The staged training workflow of \model: \ding{202} Training the \effectivemodel (BERT) passage module \ding{203} Training the full model on a document collection without selection (all available passages of a document are scored with \effectivemodel). \ding{204} The \efficientmodel (CK) selection module is now trained via knowledge distillation using the \effectivemodel (BERT) scores as labels.}
    \label{fig:training-workflow-combined}
    \vspace{-0.2cm}
\end{figure*}

\subsection{The \model Architecture}
\model is designed based on a cascade architecture within the documents. For each query-document pair, the idea is to select a few passages from the document using an efficient model and then produce the retrieval score for the document by exploiting a more expensive and effective model on the selected passages. The high-level architecture of \model is presented in \figurename~\ref{fig:model-workflow-combined}.

\model takes a query $q$ and a document $d$ as input. It first divides the document into multiple units of partially overlapping windows of size $w$ with an overlapping factor of $o$, where $o < w$. This results in $\ceil{dl/w}$ passages as follows:

\begin{equation}
\begin{aligned}
P &= [(d_{1-o:w+o})(d_{w-o:2w+o})(d_{2w-o:3w+o});...]
\end{aligned} 
\end{equation}

Each passages contains $w+2o$ tokens with exactly $2o+1$ tokens in common with their previous and next passages respectively. The first and the last passages are padded. A key practical impact of padding the windows (as opposed to padding the document) is the possibility to compact batched tensor representations and skip padding-only windows entirely for batches that contain documents of different lengths. 

\model uses an efficient model to score each passage $p \in P$ with respect to the query. This model is called \textbf{\efficientmodel}. To cascade the scoring decision and discard a large number of passages, \model selects the top $k$ passages with the highest scores produced by the \efficientmodel model, as follows:
\begin{equation}
    \hat{P} = \argmax_{\hat{P} \subseteq P, |\hat{P}| = k} \sum_{p \in \hat{P}}{\text{\efficientmodel}(q, p)}
\end{equation}

It is important to keep the size of $\hat{P}$ as small as possible to use the efficiency advantages brought by the cascading approach. 

The selected passages are then scored by a more expensive and effective ranking model, called \textbf{\effectivemodel}:
\begin{equation}
S_\text{\effectivemodel}  = \text{\effectivemodel}(q, p) \: \big| \: p \in \hat{P}
\end{equation}

Note that the token embedding parameters are shared between the \efficientmodel and \effectivemodel models. We compute the document relevance score using a weighted linear interpolation. In other words, we feed the top $l$ sorted \effectivemodel scores to a fully-connected layer to produce the document relevance score as follows:
\begin{equation}
\text{\model}(q, d) = \text{top}_l(S_\text{\effectivemodel}) * W_{PS}
\end{equation}
where $W_{PS}$ is a $l \times 1$ weight matrix for linear interpolation of the passage scores.

%\newpage
\subsection{The \model Optimization}
\label{sec:training}
The \model framework consists of multiple non-differentiable operations (e.g., passage selection) that makes it difficult to use gradient descent-based methods for end-to-end optimization. Therefore, we split the \model optimization to three steps with different objectives, as shown in \figurename~\ref{fig:training-workflow-combined}. These three steps include: (1) optimizing the \effectivemodel model for passage ranking, (2) extending the \effectivemodel optimization to full document ranking, and (3) optimizing the \efficientmodel model for passage selection using knowledge distillation.
Each step completes with early stopping based on the performance on a held-out validation set and the best model checkpoint is used as the initialization for the following step(s). The first step involves training the passage ranking model.

\paragraph{\textbf{Step I: Optimizing \effectivemodel for Passage Ranking}}
The first training step is to train the \effectivemodel model on a passage collection. To this aim, we adopt the pairwise ranking loss function used in RankNet~\cite{burges2010ranknet}. In more detail, for a given pair of negative and positive passages $p^-$ and $p^+$ for the query $q$ in the training set, we use a binary cross-entropy loss function for pairwise passage ranking optimization as follows:
\begin{equation}
\mathcal{L}_\text{Pas.}(q , p^{+} , p^{-}) = -\log \sigma(\text{\effectivemodel}(q, p^+) - \text{\effectivemodel}(q, p^-))
\end{equation}
where $\sigma(\cdot)$ is the sigmoid function. 

This step prepares the \effectivemodel model for the document retrieval task. Such pre-training has been successfully employed in recent models, such as PARADE~\cite{li2020parade}. The parallel MSMARCO passage and document collections make this pre-training possible, albeit it remains optional if passage relevance is not available, as the BERT module is also trained in the next step.

\paragraph{\textbf{Step II: Extending the \effectivemodel Optimization to Full-Document Ranking}}
Optimizing the model for passage ranking is not sufficient for document retrieval, mainly because of the following two reasons: First, the passage aggregation parameters (i.e., $W_{PS}$) need to be optimized; Second, the passage and document collections may exhibit different assumptions on what constitutes relevance. Therefore, in the second optimization step, we train the \effectivemodel model in addition to the passage aggregation layer using a \textit{full} document ranking setting, in which there is no passage selection and all passages are scored and the top $l$ passages are chosen for the aggregation layer (i.e., $W_{PS}$). We initialize the \effectivemodel parameters with the best checkpoint obtained from early stopping in the previous optimization step. We again use the binary cross-entropy loss function, this time for a query and a pair of positive and negative documents.

This optimization step further fine-tunes the \effectivemodel parameters and learns the $W_{PS}$ parameters.

\paragraph{\textbf{Step III: Optimizing \efficientmodel for Passage Selection using Knowledge Distillation}}
The last two optimization steps give us an effective document ranking model that runs 
\effectivemodel on every passage in the document and aggregates the scores. In this step, we optimize the \efficientmodel parameters. Given the fact that the goal of \efficientmodel is to select passages to be consumed by \effectivemodel in a cascade setting, we use knowledge distillation for training the \efficientmodel model. In other words, we optimize the \efficientmodel parameters such that it mimics the \effectivemodel behavior using a teacher-student paradigm, where \efficientmodel and \effectivemodel play the roles of student and teacher, respectively. Therefore, the output of \effectivemodel provides labels for the \efficientmodel model. A similar idea has been employed in the weak supervision literature~\cite{dehghani2017fidelity}. Formally, the loss function for this optimization step is defined as:
\begin{equation}
\mathcal{L}_\text{Selection}(q , d) = \mathcal{L}_\text{KD}\big ( [\text{\effectivemodel}(q, p)], [\text{\efficientmodel}(q, p)] \big)  \: \big| \: p \in P
\end{equation} 
where $P$ denotes all the passages in the document $d$ and $\mathcal{L}_\text{KD}$ denotes a knowledge distillation loss function. The function $\mathcal{L}_\text{KD}$ is responsible for computing the average across passages. A unique feature of our distillation approach is that the teacher signals created by \effectivemodel are unsupervised. It is important to train the less capable \efficientmodel on the exact distribution it is later used. There are no passage-level labels for all MSMARCO documents we could use to train the \efficientmodel, therefore the \effectivemodel is the only training signal source.

In our experiments, we study multiple loss functions for knowledge distillation. They include distribution losses, such as mean square error (MSE) and cross entropy, and in-document passage ranking loss functions, such as nDCG2 introduced as part of the LambdaLoss framework~\cite{Wang2018lamdaloss}. The nDCG2 loss function is a gain-based loss to tightly bind the loss to NDCG-like metrics. For the exact formulation we refer to~\citet{Wang2018lamdaloss}. In the nDCG2 loss, we assign gain values only to the top $k$ passages sorted by \effectivemodel and all other passages receive no gain. 
The nDCG2 loss, focusing on moving the correct $k$ passages in the top positions is a great fit for our problem: The \efficientmodel is only used for pruning or filtering, which means that the ordering inside and outside the top-$k$ set does not matter. The only thing that matters is to find the right set of $k$ passages, as the \effectivemodel then creates our final fine-grained scores for each of the $k$ passages. This is a crucial difference in our knowledge distillation approach to concurrent works, which try to fit every decision from the more powerful ranking model to a smaller model \cite{li2020parade}. 
Not surprisingly, we find that using the nDCG2 ranking loss outperforms other loss functions, as discussed in Section \ref{sec:distillation}. 

\subsection{The \model Implementation}
In this section, we describe the implementation details for the \efficientmodel and \effectivemodel models used in our experiments.

\paragraph{\textbf{\efficientmodel: The CK Model}}
The \efficientmodel model is the first model in our cascaded architecture and is expected to be extremely efficient. In our experiments, we use CK, an efficient variation of the Conv-KNRM model~\cite{Dai2018} that combines convolutional neural networks (CNNs) with the kernel-pooling approach of~\citet{Xiong2017}. Unlike Conv-KNRM that uses multiple convolutional layers with different window sizes for soft-matching of n-grams in the query and document, the CK model uses a single convolutional layer to provide local contextualization to the passage representations without the quadratic time or memory complexity required by Transformer models. In more detail, the CK model transforms the query and passage representations using a CNN layer and uses the cosine function to compute their similarities, which are then activated by Gaussian kernels with different distribution parameters:
\begin{equation}
\begin{aligned}
    K^{k}_{i,j} &= \exp \left(-\frac{\left(\cos(\cnn(q_i),\cnn(p_j))-\mu_{k}\right)^{2}}{2 \sigma^{2}}\right) \\
\end{aligned}
\end{equation}
where $q_i$ and $p_j$ respectively denote the $i$\textsuperscript{th} token in the query and the $j$\textsuperscript{th} token in the passage. $\mu_k$ and $\sigma$ are the Gaussian kernel parameters.
Each kernel represents a feature extractor, which is followed by a pooling layer that sums up the individual activations, first by the passage dimension $j$ and then log-activated by the query token dimension $i$. Each kernel result is weighted and summed with a single linear layer ($W_k$) as follows:
\begin{equation}
\begin{aligned}
    \ck(q, p) &= \bigg(\sum_{i=1}^{|q|} \log\left( \sum_{j=1}^{|p|} K^{k}_{i,j} \right) \bigg) * W_K
\end{aligned}
\end{equation}

\paragraph{\textbf{\effectivemodel: The BERT Ranking Model}}
Large-scale pre-trained language models, such as BERT~\cite{devlin2018bert}, have led to state-of-the-art results in a number of tasks, including passage retrieval~\cite{nogueira2019passage}. The BERT passage ranking model takes sequences representing a query $q$ and a passage $p$ and concatenates them using a separation token. The obtained BERT representation for the first token of the query-passage pair (i.e., the \texttt{[CLS]} token) is then fed to a fully-connected layer ($W_s$) to produce the ranking score:
\begin{equation}
\begin{aligned}
    \bert(q, p) &= \bert_{\texttt{[CLS]}}(\texttt{[CLS]}; q; \texttt{[SEP]}; p) * W_s
\end{aligned}
\end{equation}
where $;$ denotes the concatenation operation.

\section{Experiment Design}
In this section, we describe our experiment setup. We implemented our models using the HuggingFace Transformer library \cite{wolf2019huggingface} and PyTorch~\cite{pytorch2017}. We employed PyTorch' mixed precision training and inference throughout our experiments for efficiency. In our experiments, we re-rank the documents retrieved by BM25 implemented in Anserini \cite{Yang2017}. The query latency measurements are conducted on the same single TITAN RTX GPU. 

\subsection{Document Collections and Query Sets}

For our first passage-training step we utilize the MSMARCO-Passage collection and training data released by \citet{msmarco16}. We follow the setup of \citet{hofstaetter2020_crossarchitecture_kd} for training the BERT passage ranker. For passage results of the BERT ranking model we refer to the previous work. In all datasets, we limit the query length at $30$ tokens to remove only very few outliers, and re-rank 100 documents from BM25 candidates. 

We use two query sets for evaluating our models as follows:

\paragraph{\textbf{TREC DL 2019 and MS MARCO benchmarks}}
We use the 2019 TREC Deep Learning Track Document Collection \cite{trec2019overview} that contains 3.2 million documents with a mean document length of 1,600 words and the 80$^\text{th}$ percentile at 1,900 words. We aim to include them with a 2,000 token limit on our experiments. We selected 5,000 queries from the training set as validation set for early stopping and removed those queries from the training data. We use the following two query sets in our evaluation:
\begin{itemize}
    \item TREC DL 2019: 43 queries used in the 2019 TREC Deep Learning Track for document ranking task. A proper pooling methodology was used to create a complete set of relevance judgments for these topics~\cite{trec2019overview}. For evaluation metrics that require binary relevance labels (i.e., MAP and MRR), we use a binarization point of 2.
    \item MS MARCO: 5,193 queries sampled from the Bing query logs contained in the MS MARCO Development set~\cite{msmarco16}. This query set is larger than the previous one, but suffers from incomplete relevance judgments. 
\end{itemize}

\begin{table*}[t!]
    \centering
    \caption{Effectiveness results for TREC-DL'19 and MSMARCO DEV query sets. Our aim is to hold the effectiveness of an All-BERT configuration with a cascaded efficiency improvement. * is a stat.sig. difference to All-BERT (2K); paired t-test ($p < 0.05$).}
    \label{tab:all_results}
    %\vspace{-0.3cm}
    \setlength\tabcolsep{3.8pt}
    %\small
    \begin{tabular}{clllr!{\color{lightgray}\vrule}rrr!{\color{lightgray}\vrule}rrr}
       \toprule
       &\multirow{2}{*}{\textbf{Model}}& \multirow{2}{*}{\textbf{Cascade}}& \textbf{\# BERT} &
       \textbf{Doc.}   &
       \multicolumn{3}{c!{\color{lightgray}\vrule}}{\textbf{TREC DL 2019}}&
       \multicolumn{3}{c}{\textbf{MSMARCO DEV }}\\
       &&&\textbf{Scored}& \textbf{Length}& nDCG@10 & MRR@10 & MAP@100 & nDCG@10 & MRR@10 & MAP@100 \\
        \midrule
        \multicolumn{7}{l}{\textbf{Baselines}} \\
        \textcolor{gray}{1} & BM25     & -- & -- & -- &   0.488 & 0.661 & 0.292 & 0.311 & 0.252 & 0.265 \\
        \textcolor{gray}{2} & TKL \cite{Hofstaetter2020_sigir} & -- & -- & 2K  & 0.634 & 0.795 & 0.332 & 0.403 & 0.338 & 0.345 \\
        \textcolor{gray}{3} & PARADE$_\text{Max-Pool}$ \cite{li2020parade} & -- & All & 2K  & 0.666 & 0.807 & 0.343 & 0.445 & 0.378 & 0.385  \\
        \textcolor{gray}{4} & PARADE$_\text{TF}$ \cite{li2020parade} & -- & All & 2K  & 0.680 & 0.820 & \textbf{0.375} & 0.446 & 0.382 & 0.387  \\
        \arrayrulecolor{lightgray}

        \midrule
        \multicolumn{7}{l}{\textbf{Ours}} \\
         \textcolor{gray}{5} & \multirow{7}{*}{\model} & -- & All & 512 &   0.667 & 0.815 & 0.348 & *0.440 & *0.374 & *0.383 \\
         \textcolor{gray}{6} &   & -- & All & 2K  &   \textbf{0.688} & 0.867 & 0.364 & \textbf{0.450} & \textbf{0.384} & \textbf{0.390} \\

         \cmidrule{3-11}

        %\midrule
          \textcolor{gray}{7} &  & Static First & 3         & 2K & 0.638 & 0.785 & 0.309 & *0.394 & *0.330 & *0.338 \\
          \textcolor{gray}{8} &  & Static Top-TF & 3        & 2K & *0.624 & 0.778 & 0.324 & *0.393 & *0.329 & *0.337 \\

         \cmidrule{3-11}
         \textcolor{gray}{9} &  & CK (nDCG2)    & 3     & 2K & 0.671 & 0.876 & 0.361 & 0.438 & 0.375 & 0.380 \\

         \textcolor{gray}{10} &   & CK (nDCG2)     & 4    & 2K & \textbf{0.688} & \textbf{0.916} & {0.365} & 0.446 & 0.380 & 0.387 \\

         %\arrayrulecolor{lightgray}
         %\midrule

        \arrayrulecolor{black}
        \bottomrule
    \end{tabular}
    %\vspace{-0.3cm}
\end{table*}

\subsection{Training Configuration}

We use Adam optimizer \cite{kingma2014adam} with a learning rate of $7*10^{-6}$ for all BERT layers. CK layers contain much fewer and randomly initialized parameters and therefore are trained with a higher learning rate of $10^{-5}$. We employ early stopping, based on the best nDCG@10 value on the validation set. 

\subsection{Model Parameters}

We use a 6-layer DistilBERT \cite{sanh2019distilbert} knowledge distilled from BERT-Base on the MSMARCO-Passage collection \cite{hofstaetter2020_crossarchitecture_kd} as initialization for our document training. We chose DistilBERT over BERT-Base, as it has been shown to provide a close lower bound on the results at half the runtime for training and testing \cite{sanh2019distilbert,hofstaetter2020_crossarchitecture_kd}. In general our approach is agnostic to the BERT-variant used, when using a language model with more layers and dimensions, the relative improvements of our cascade become stronger, at the cost of higher training time and GPU memory. 

For the passage windows we set a base size $w$ of 50 and an overlap of $7$ for a total window size of $64$. In a pilot study, we confirmed that a larger window size does not improve the All-BERT effectiveness results.
For the BERT passage aggregation, we use the top 3 BERT scores to form the final document score, independent of the cascade selection count. This allows us to base all selection models on the same All-BERT instance.

We set the token context size for CK to 3 and evaluate two different CK dimensions, first a full 768 channel convolution corresponding to the dimensions of the BERT embeddings and second a smaller convolution with a projection to 384 dimensions before the convolution and another reduction in the convolution output dimension to 128 per term, which we refer to as CKS.

\section{Results}
In this section, we address the research questions raised in Section~\ref{sec:intro}.

\subsection{RQ1: Knowledge Distillation and Effectiveness Study}
\label{sec:distillation}

Our first research question centers around training techniques and we investigate: 
\RQone

To address this research question, we compare the proposed method to a number of strong retrieval baselines, including BM25, TKL~\cite{Hofstaetter2020_sigir}, and PARADE~\cite{li2020parade}. TKL is a non-BERT local self-attention ranking model with kernel-pooling; PARADE\textsubscript{MAX-Pool} is very close to our All-BERT baseline, in that it scores every passage with a BERT ranker and aggregates passage representations in a lightweight layer; PARADE\textsubscript{TF} uses an additional Transformer block to aggregate passage representations. The results are reported in \tablename~\ref{tab:all_results}. The first observation on the All-BERT setting of \model, without cascading, is the strong difference in effectiveness across collections between 512 document tokens and 2K tokens (Line 5 vs. 6). The All-BERT 2K model (Line 6) outperforms all previously published single model re-ranking baselines on the TREC DL 2019 dataset, except for PARADE\textsubscript{TF} (Line 4) on MAP . We see how training and evaluating with longer document input improves the effectiveness. This is also a strong argument for our cascading approach that makes it possible to process long inputs. We chose the All-BERT 2K setting (Line 6) as the base model for all following results. 

Furthermore we use static cascade selectors as baselines, that use BERT scoring after static selections. One selects the first 3 passages by position, to have a direct comparison to CK selections (Line 7). Another option we benchmark is the use of raw term frequency matches, where we take the three passages with the highest frequency of direct term matches without semantic or trained matching (Line 8). Both approaches fail to improve the scoring over the baselines or our CK selection model. The term frequency selection even underperforms the first positions selector.

Our main \model configuration (Lines 9 \& 10) with a knowledge distilled CK shows strong results, which are not statistically different to their base model of All-BERT (2K, Line 6). Across both query sets, the select 3 (Line 9) is already close to the reachable results and on select 4 (Line 10) we obtain better MRR and MAP results on TREC-DL'19, even though they are not significantly different.

\begin{table}[t!]
    \centering
    \caption{Impact of knowledge distillation with measures using a cutoff at 10. * indicates stat.sig. difference; paired t-test ($p < 0.05$).}
    \label{tab:kd_ablaition}
    \vspace{-0.3cm}
    \setlength\tabcolsep{6pt}
    %\small
    \begin{tabular}{ll!{\color{lightgray}\vrule}rr!{\color{lightgray}\vrule}rr}
       \toprule
       \textbf{Scoring}& \multirow{2}{*}{\textbf{Training}}&
       
       \multicolumn{2}{c!{\color{lightgray}\vrule}}{\textbf{TREC-DL'19}}&
       \multicolumn{2}{c}{\textbf{MSMARCO}}\\
       \textbf{Model} &&  nDCG & MRR & nDCG & MRR  \\
       \midrule
        \arrayrulecolor{lightgray}
         CK & Standalone   & 0.551 & 0.677  & 0.353 & 0.287  \\
         CK & BERT-KD      & *\textbf{0.595} & \textbf{0.749}  & *\textbf{0.363} & *\textbf{0.299}  \\
        \arrayrulecolor{black}
        \bottomrule
    \end{tabular}
    \vspace{-0.3cm}
\end{table}

\begin{table}[t!]
    \centering
    \caption{Knowledge distillation loss study of the \model cascade training with measures using a cutoff at 10. Stat.sig. is indicated with the superscript to the underlined character; paired t-test ($p < 0.05$).}
    \label{tab:loss_func_res}
    \vspace{-0.3cm}
    \setlength\tabcolsep{5.5pt}
    %\small
    \begin{tabular}{cl!{\color{lightgray}\vrule}rr!{\color{lightgray}\vrule}rr}
       \toprule
       \textbf{\# BERT}& \multirow{2}{*}{\textbf{KD-Loss}}&
       
       \multicolumn{2}{c!{\color{lightgray}\vrule}}{\textbf{TREC-DL'19}}&
       \multicolumn{2}{c}{\textbf{MSMARCO}}\\
       \textbf{Scored} &&  nDCG & MRR & nDCG & MRR \\
       \midrule
        \arrayrulecolor{lightgray}
         %\midrule
         \multirow{3}{*}{3} & MSE   & 0.664 & 0.816  & 0.426 & 0.362  \\ 
          & Cross Entropy           & 0.667 & 0.851  & 0.437 & 0.373  \\ 
          & nDCG2                   & \textbf{0.671} & \textbf{0.876}  & \textbf{0.438} & \textbf{0.375}  \\ 
         \midrule
         \multirow{3}{*}{4} & \underline{M}SE    & 0.683 & 0.870 & 0.437 & 0.374 \\ 
          & Cross Entropy            & 0.675 & 0.889 & $^{m}$\textbf{0.446} & $^{m}$\textbf{0.381} \\ 
          & nDCG2                    & \textbf{0.688} & \textbf{0.916} & $^{m}$\textbf{0.446} & $^{m}$0.380 \\

        \arrayrulecolor{black}
        \bottomrule
    \end{tabular}
    %\vspace{-0.3cm}
\end{table}

We further compare two training strategies for the efficient CK model: (1) knowledge distillation as described in Section~\ref{sec:training}, and (2) standalone training on relevance judgments. The results are reported in Table \ref{tab:kd_ablaition}. We observe that the knowledge distilled CK model leads to significantly higher performance on both TREC DL 2019 and MS MARCO datasets in comparison to the standalone training. The improvements on the TREC DL 2019 query set are much larger. As expected, CK alone shows substantially lower effectiveness compared to BERT. Overall we show that even though alone CK is ineffective, combined with BERT in the \model model produces very effective and efficient results. 

We extend our analysis by inspecting different knowledge distillation losses and their effect on the full cascaded setting in \tablename~\ref{tab:loss_func_res}. We probe knowledge distillation with an MSE, Cross Entropy and the LambdaLoss nDCG2 losses for two different cascade selection numbers, three and four. 
When four passages are selected, all three knowledge distillation losses perform on par with the All-BERT model. Overall using nDCG2 as the knowledge distillation loss outperforms the other two loss functions and is the most stable across the two evaluated query sets. Therefore, in the following experiments, we only focus on nDCG2 as the default knowledge distillation loss.

\subsection{RQ2: Efficiency-Effectiveness Analysis}

\begin{figure}[t]
  \centering
  \begin{subfigure}[t]{0.49\textwidth}
    \centering
    \includegraphics[width=.9\textwidth,clip, trim=0.2cm 0.2cm 0.2cm 0.2cm]{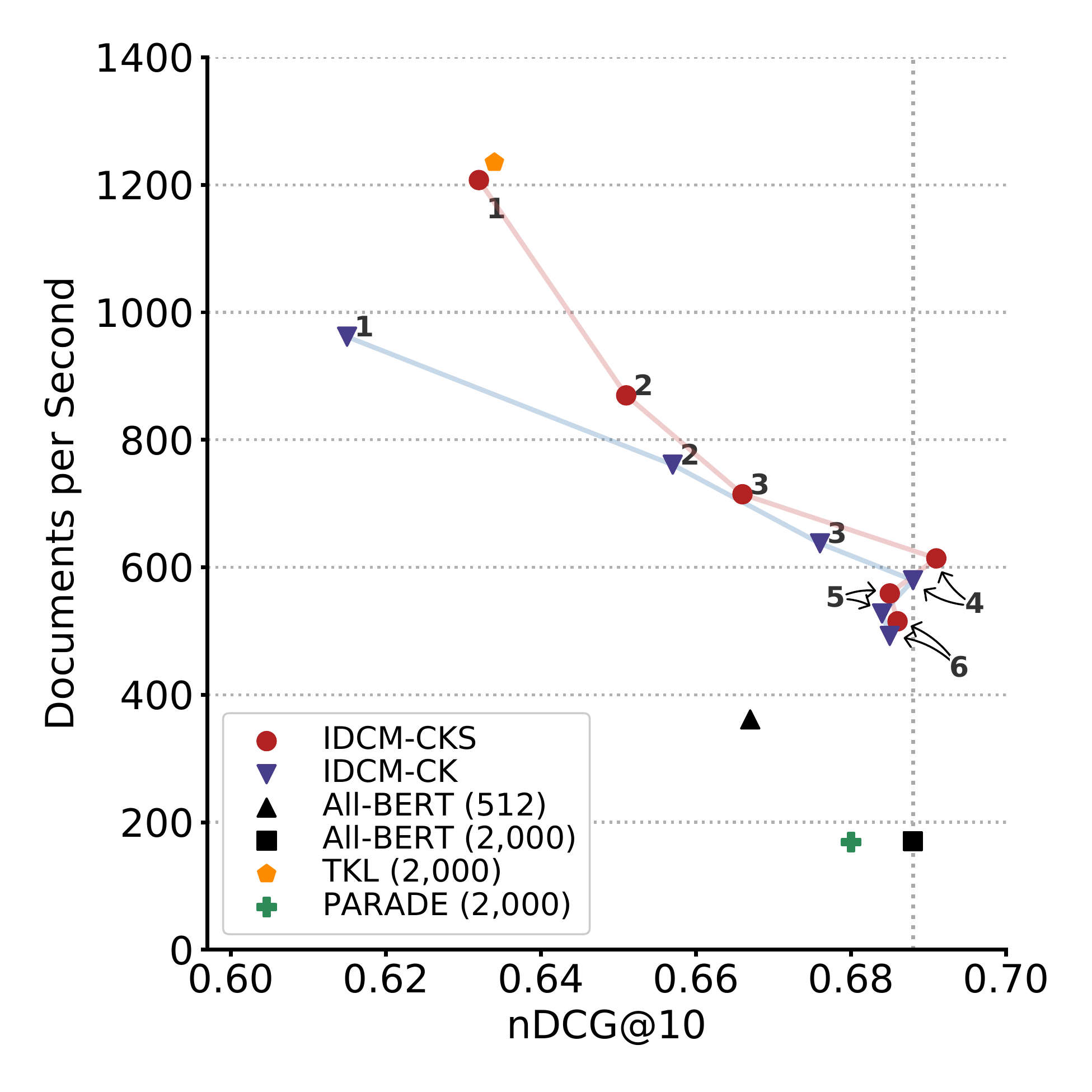}
    \vspace{-0.4cm}
    \caption{Throughput and nDCG@10 results on TREC DL 2019.}
    \label{fig:throughput-ndcg-trec}
  \end{subfigure}
  \vspace{0.4cm}
%   \hspace{1em}
  \begin{subfigure}[t]{0.49\textwidth}
    \centering
    \includegraphics[width=.9\textwidth,clip, trim=0.2cm 0.2cm 0.2cm 0.2cm]{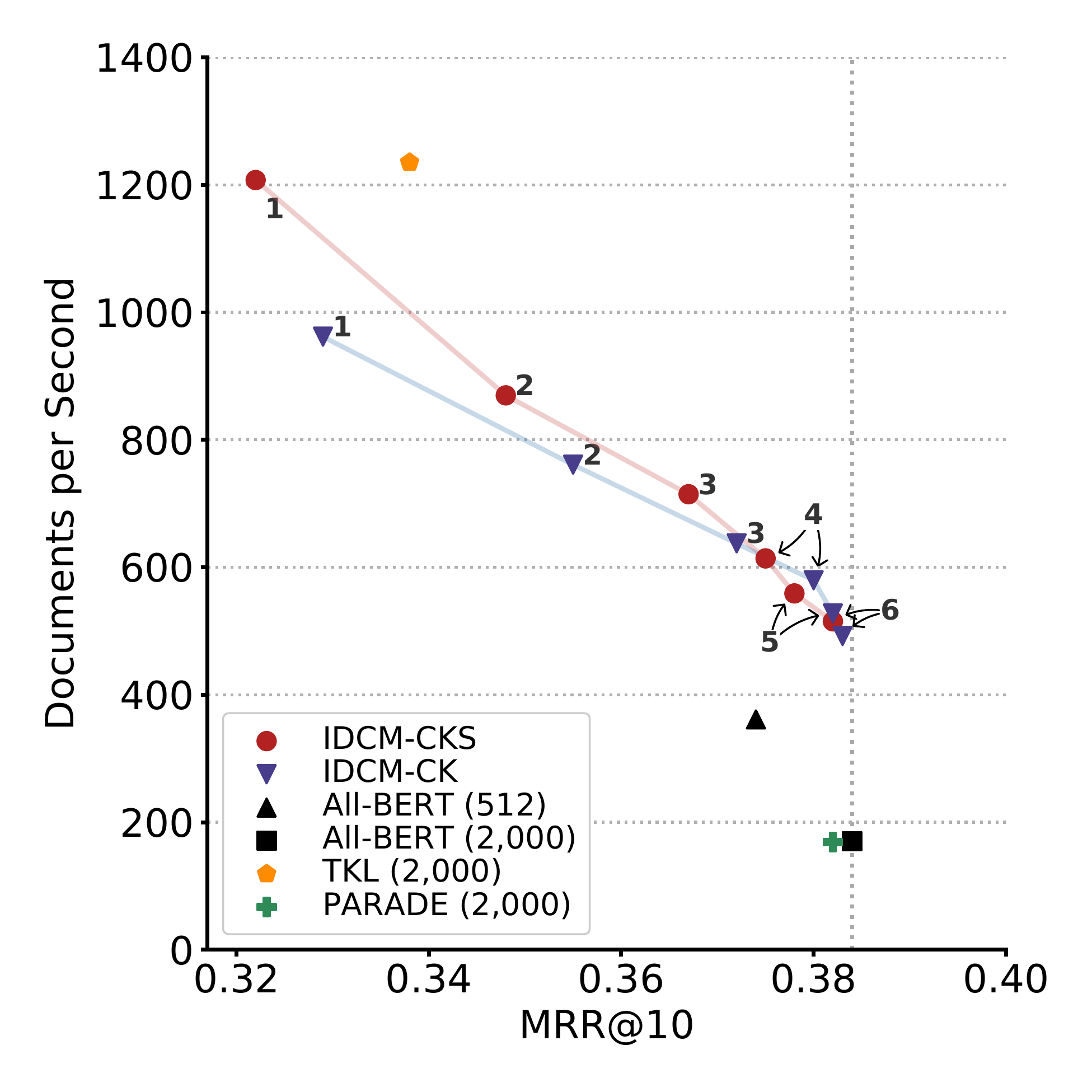}
    \vspace{-0.4cm}
    \caption{Throughput and MRR@10 results on MS MARCO Dev.}
    \label{fig:throughput-mrr-msmarco}
  \end{subfigure}
    \vspace{-0.4cm}
  \caption{Throughput and ranking effectiveness trade-off results. The vertical line shows the achievable effectiveness, for all \model models based on All-BERT (2,000). The number next to the marker indicates the selection count.}
  \vspace{-0.4cm}
%   \label{fig:x}
\end{figure}

\begin{figure*}
    \centering
    %trim={<left> <lower> <right> <upper>}
    \includegraphics[width=0.85\textwidth,clip, trim=0.2cm 0.2cm 0.2cm 0.2cm]{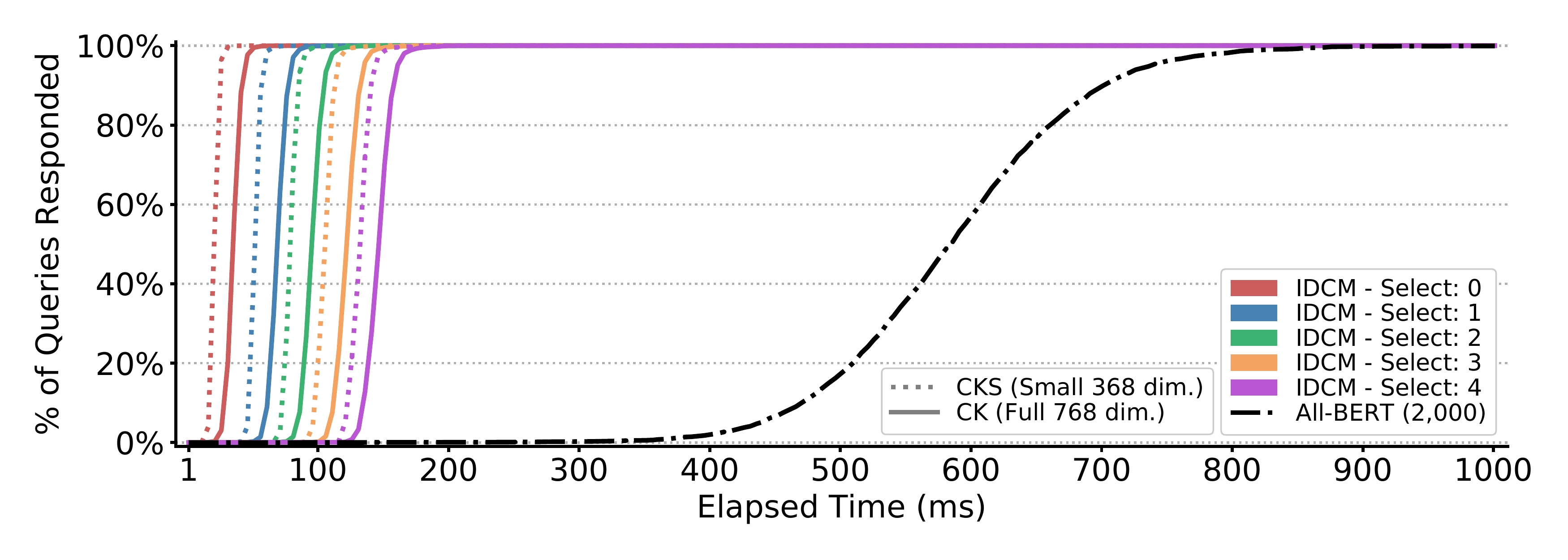}
    \vspace{-0.5cm}
    \caption{Fraction of queries that can be answered in the given time-frame for re-ranking 100 documents with up to 2,000 tokens on MSMARCO. Select 0 means only CK timing without BERT cascading.}
    \label{fig:query-latency}
    \vspace{-0.4cm}
\end{figure*}

To study our next research question we turn to the viewpoint of both efficiency and effectiveness of different \model model configurations to answer:

\RQtwo

The selection variants of \model are all based on the All-BERT setting and therefore this model instance sets our potential for cascaded effectiveness results, as we do not train the BERT module during selection training. In Figures \ref{fig:throughput-ndcg-trec} and \ref{fig:throughput-mrr-msmarco} we compare the model throughput of documents per second (y-axes) with their effectiveness (x-axes). Akin to a precision-recall plot, the best result would be situated in the upper right corner.

We evaluated \model's selection parameter -- the number of passages scored with the costly BERT model -- $k$ from 1 to 6 using the full convolution CK and a smaller dimensional CKS setting. We find that selecting too few passages reduces the effectiveness strongly, however starting with a selection of 4 or more passages, \model results are very similar to All-BERT results, while providing a much higher throughput. On the nDCG@10 metric of TREC-DL'19 in Figure \ref{fig:throughput-ndcg-trec} \model reaches the All-BERT effectiveness starting with 4 selected passages. In Figure \ref{fig:throughput-mrr-msmarco} the \model setting is already close to the reachable effectiveness, and taking 5 and 6 passages close the gap to All-BERT further, to a point of almost no differences.

A simple efficiency improvement is to use a lower document length on All-BERT passage scoring, such as limiting the document to the first 512 tokens. We find that this works both more slowly and less effectively than \model. 

It is important to note, that in all our experiments presented here we utilize the 6-layer DistilBERT encoder. When we compare with related work, which commonly uses larger BERT-style models, such as the original BERT-Base or BERT-Large, we show even better efficiency improvements. A 12-layer BERT-Base model in the All-BERT (2,000) configuration can only process 85 documents per second, and the 24-layer BERT-large only manages to score 30 documents per second. Applying our \model technique to these larger encoders, brings even larger performance improvements.

In summary, with this analysis we show how \model can be as effective as an All-BERT model, while maintaining a four times higher throughput. In addition, users of the \model model have the option to trade-off effectiveness for more efficiency, along a clear curve.

\begin{figure}
    \centering
    %trim={<left> <lower> <right> <upper>}
    \includegraphics[width=0.4\textwidth,clip, trim=0.6cm 0.2cm 0.2cm 0.2cm]{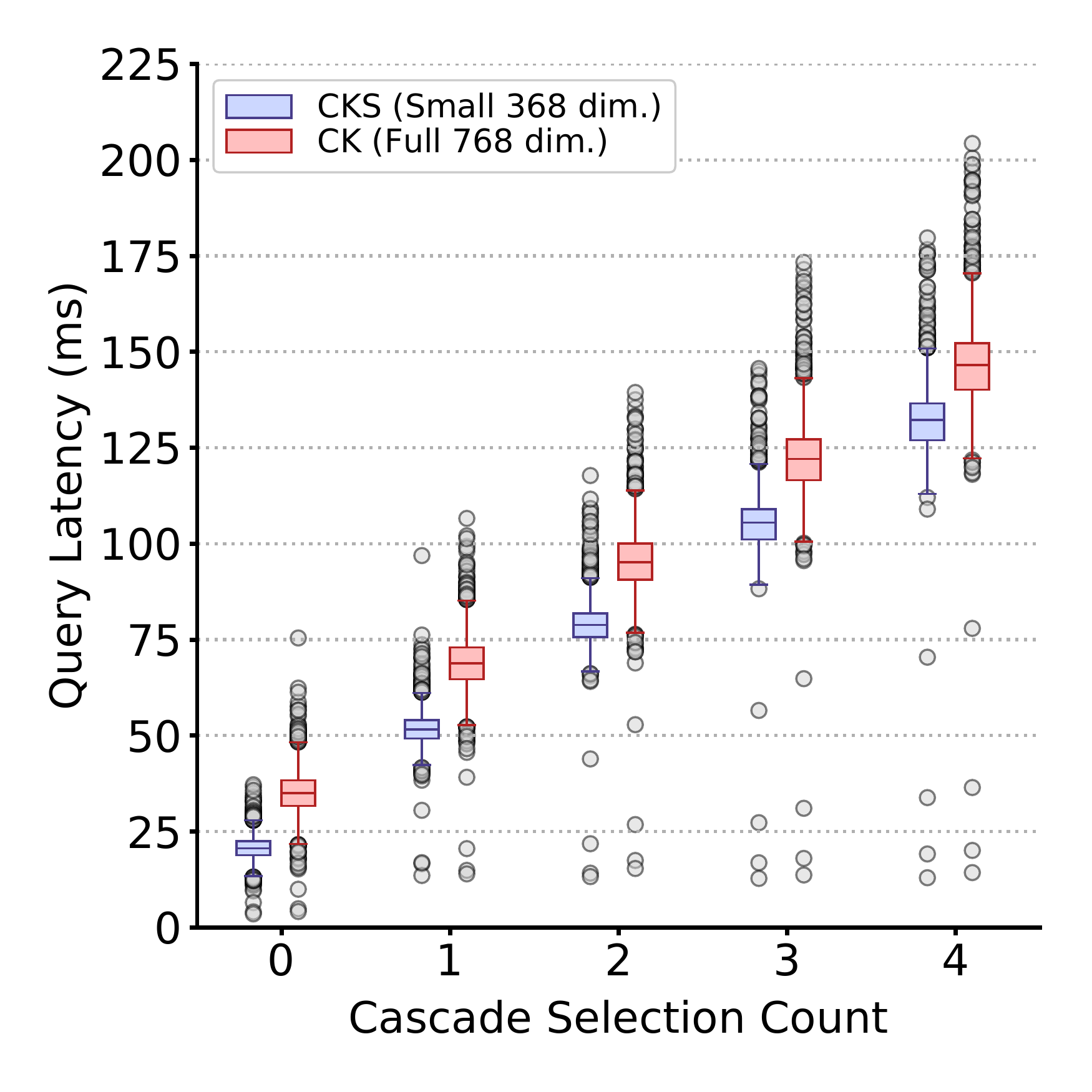}
    \vspace{-0.3cm}
    \caption{Query latency for different \model cascade selection configurations for re-ranking 100 documents with up to 2,000 tokens on MSMARCO. Selection 0 only measures the time for the CK module without routing passages to BERT.}
    \label{fig:query-latency-boxplot}
    \vspace{-0.3cm}
\end{figure}

\subsection{RQ3: Query Latency Analysis}

The mean or median aggregations of query latencies hide crucial information about the tail of queries that require more processing time. In the case of neural ranking models this tail is heavily dependent on the total document length of all documents in a batch. 
We now study efficiency in detail with:

\RQthree

In Figure \ref{fig:query-latency} we plot the fraction of queries (y-axis) that can be processed by the neural models in the given time (x-axis). The All-BERT baseline (dash-dotted black line) for documents up to 2,000 tokens has a large range in the time required to re-rank documents. The All-BERT setting of \model already is a strong baseline for query latency, as we compact passage representations, to skip padding-only passages. However, it still requires BERT to run on up to 40 passages. Our \model model configurations with two CK sizes (dotted and full line) show a much lower variance and overall faster query response. Naturally, the more passages we select for cascading to the BERT module, the longer \model takes to compute.

Now, we turn to a more focused query latency view of the different \model configurations with boxplots of the query latency distributions in Figure \ref{fig:query-latency-boxplot}. Again, we report timings for a full CK (red; right side) and smaller CKS variant (blue; left side). The first entry with a selection of 0 shows only the computational cost of the CK module for 2,000 tokens without running BERT. When we compare the latency differences between 0 and 1 selection we can see that computing BERT for a single passage per document adds 25 ms to the median latency. This shows the efficiency of CK(S): roughly 40 passages processed with CK have an equal runtime compared to a single passage that BERT requires.

\subsection{RQ4: Passage Selection Analysis}

\begin{figure}
    \centering
    %trim={<left> <lower> <right> <upper>}
    \includegraphics[width=0.4\textwidth,clip, trim=0.2cm 0.2cm 0.2cm 0.2cm]{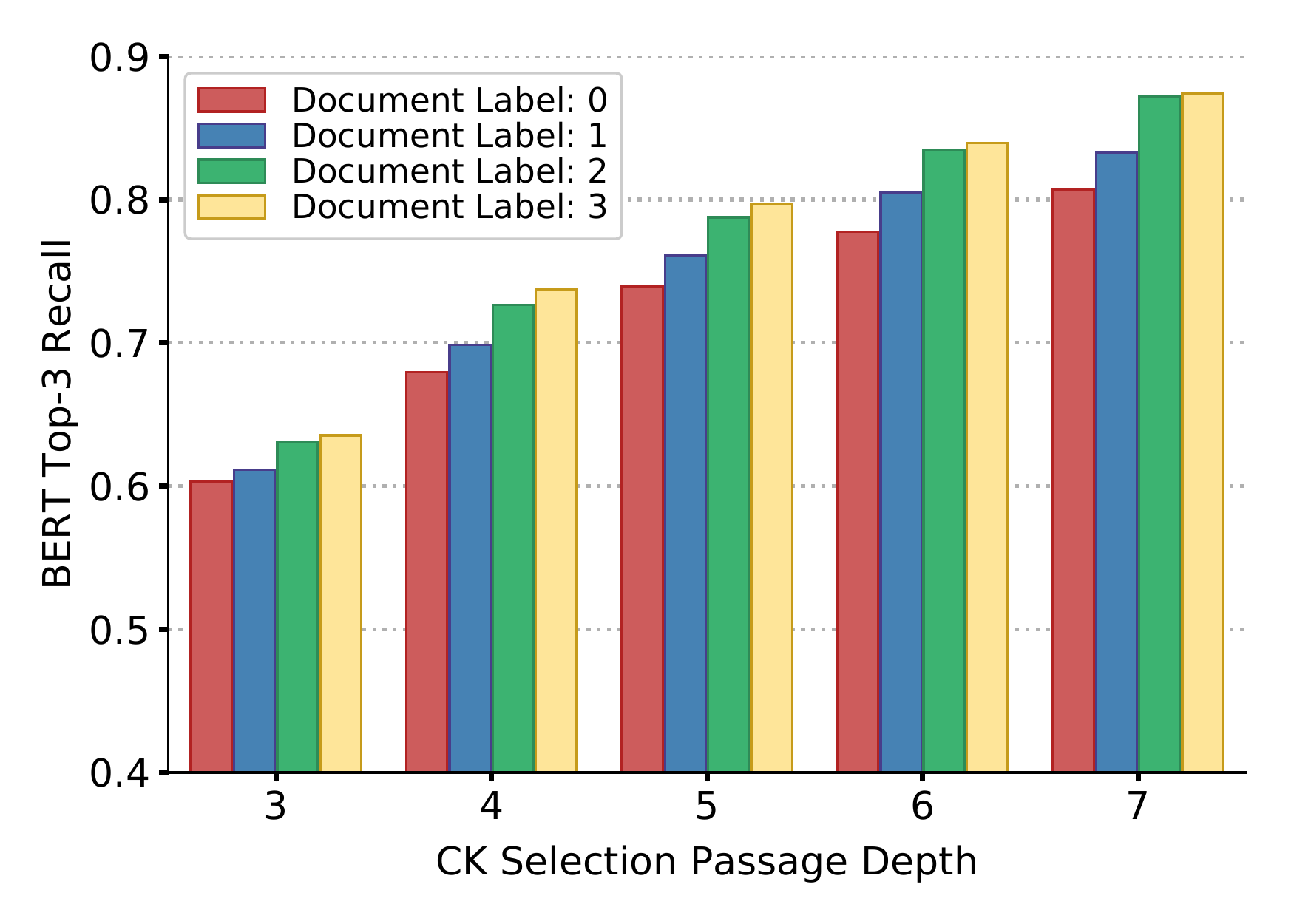}
    \vspace{-0.2cm}
    \caption{Intra-document passage selection recall of the CK selection module in comparison to the top-3 BERT selection split by document relevance grade on TREC-DL'19. The higher the label the more relevant a document.}
    \label{fig:ck-recall}
    \vspace{-0.3cm}
\end{figure}

As presented in Section \ref{sec:training} we aim to train the CK module to imitate the passage scoring of the BERT module. To understand how well the CK model is able to do just that we evaluate the intra-document passage recall of the CK scores in comparison to the top-3 BERT passages selected to form the final document score and study:

\RQfour

In Figure \ref{fig:ck-recall} we plot the CK recall for different numbers of selected CK-passages and split the reporting by document relevance grades on the TREC-DL'19 query set. We find an interesting pattern among the different relevance classes: CK is able to provide more accurate passage selections the more relevant a document is. 

A recall of 1 would guarantee the same document score, however as we showed it is not necessary for the \model model to provide very close effectiveness results to the original ranking. 

\begin{figure}
    \centering
    %trim={<left> <lower> <right> <upper>}
    \includegraphics[width=0.4\textwidth,clip, trim=0.5cm 0.2cm 0.2cm 0.2cm]{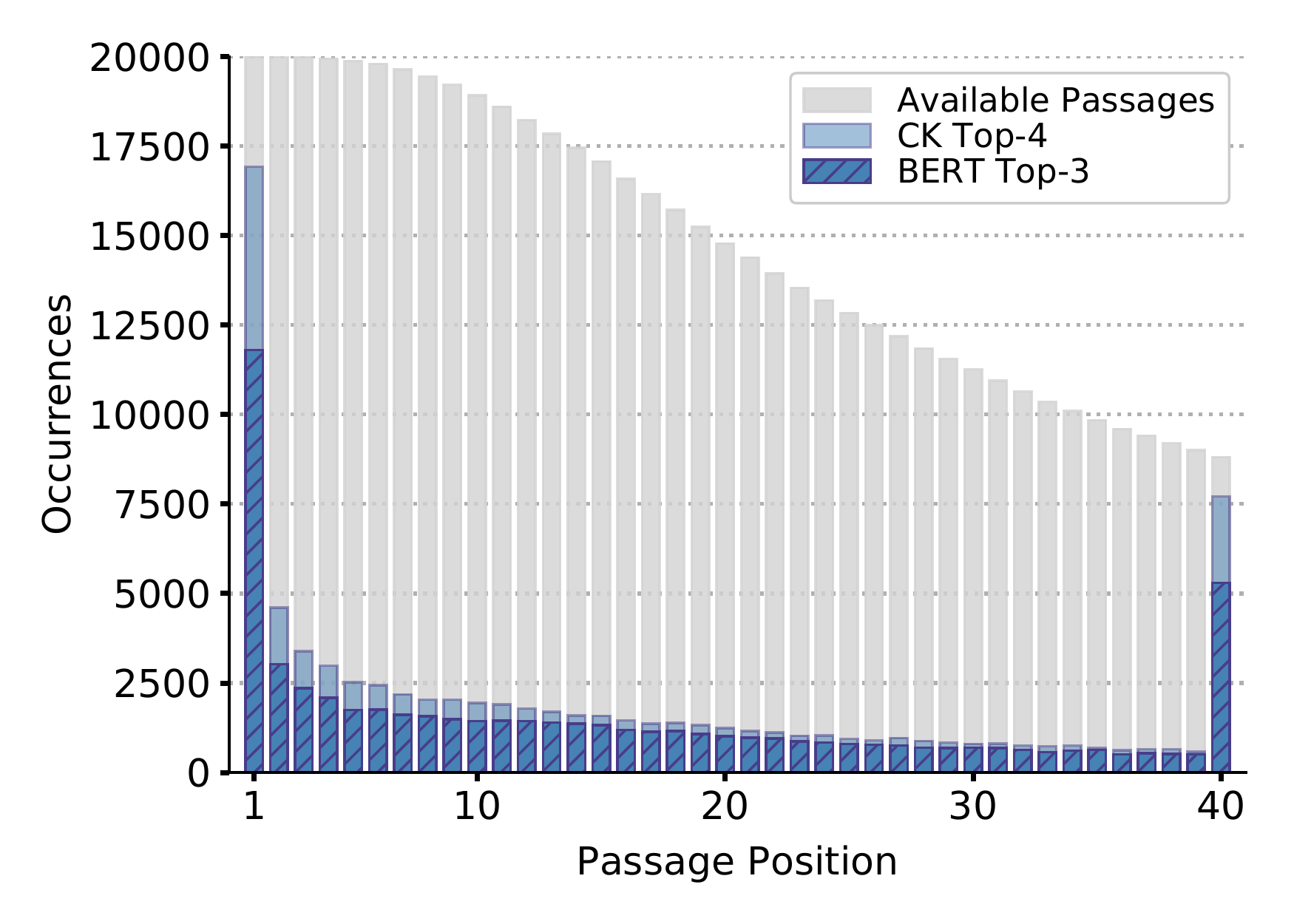}
    \vspace{-0.2cm}
    \caption{Selected Passages by the CK top-4 module and then subsequently scored by BERT for the top-3 scoring.}
    \label{fig:passage-position-analysis}
    \vspace{-0.3cm}
\end{figure}

Finally, we inspect the behavior of both BERT and CK modules with respect to the positions of the highest scoring passages. In Figure \ref{fig:passage-position-analysis} we investigate the top selections of both CK and BERT along the positions of the passages. In gray in the background are the available passages per position of the top-100 documents of the TREC-DL'19 query set, they reduce as not all documents are long enough to fit the maximum used length of 2,000 tokens. The All-BERT setting needs to compute BERT scores for all available passages, whereas in \model the selected passages in blue are the only passages score by BERT. The top-3 passages from BERT are furthermore visualized by the shaded dark blue bars. 

We see a strong focus on the first and last possible passages selected by the modules. A focus on the first passages is to be expected as the title and introduction of web pages is situated there, however the focus on the last passages is more curious. Because of our symmetric padding, the first and last passages have an empty (padded) overlap. This is the only indicator to the BERT model of passage positions in a document, the absence of 7 more tokens. It seems this is the signal the BERT model picks up on and subsequently trains the CK model to follow along. We leave deeper investigations of the behavior of BERT in different document parts for future work and conclude this section with the finding that \model learns to use all available passages including the end of a document input.

\section{Conclusion}

Applying a BERT model many times per document, once for each passage, yields state-of-the-art ranking performance. The trade-off is that inference cost is high due to the size of the model, potentially affecting both the mean and variance of query processing latency.
Typically in neural retrieval systems, one is forced to a make a clear decision between reaching efficiency or effectiveness goals. In this work we presented \model, an intra-document cascaded ranking model that provides state-of-the-art effectiveness, while at the same time improving the median query latency by more than four times compared a non-cascaded full BERT ranking model. Our two-module combination allows us to efficiently filter passages and only provide the most promising candidates to a slow but effective BERT ranker. We show how a key step in achieving the same effectiveness as a full BERT model is a knowledge distilled training using the BERT passage scores to train the more efficient selection module. Our knowledge distillation provides self-supervised teacher signals for all passages, without the need for manual annotation. Our novel distillation technique not only improves the query latency of our model in a deployment scenario, it also provides efficiency for replacing manual annotation labor and cost with a step-wise trained teacher model. In the future we plan to extend the concept of intra-document cascading for document ranking to a dynamic number of passages selected and more cascading stages.

\begin{acks}
This work was supported in part by the Center for Intelligent Information Retrieval. Any opinions, findings and conclusions or recommendations expressed in this material are those of the authors and do not necessarily reflect those of the sponsor.
\end{acks}

\balance
\bibliographystyle{ACM-Reference-Format}
\bibliography{references}

\end{document}